\pdfoutput=1
\nonstopmode

\documentclass[10pt]{article}
\usepackage{makeidx}
\usepackage{amsfonts}
\usepackage{amsmath}

\usepackage{amsfonts}
\usepackage{amsmath}
\usepackage{amssymb}
\usepackage{array}
\usepackage{amsmath}
\usepackage{amssymb}
\usepackage{url}

\usepackage{subfigure}

\usepackage{citesort}
\usepackage[numbers,round]{natbib}

\usepackage{multicol}
\makeatletter

\usepackage{ifpdf}
\ifpdf
\usepackage[%
  pdftitle={Analysis and Optimization of Pulse Dynamics in TMS},%
  pdfauthor={S. M. Goetz et al.},%
  pdfsubject={-},%
  pdfkeywords={-},%
  pdfstartview=FitH,%
  bookmarks=true,%
  bookmarksopen=true,%
  breaklinks=true,%
  colorlinks=true,%
  linkcolor=blue,anchorcolor=blue,%
  citecolor=blue,filecolor=blue,%
  menucolor=blue,pagecolor=blue,%
  urlcolor=blue]{hyperref}
\else
\usepackage[%
  breaklinks=true,%
  colorlinks=true,%
  linkcolor=blue,anchorcolor=blue,%
  citecolor=blue,filecolor=blue,%
  menucolor=blue,pagecolor=blue,%
  urlcolor=blue]{hyperref}
\fi

\usepackage{graphicx}
\usepackage{amssymb}
\usepackage{textcomp}

\begin{document}

\title{Analysis and Optimisation of Pulse Dynamics for Magnetic Stimulation}

\date{June 15, 2011}

 \author{S.M.~Goetz\footnote{Email: {\tt sgoetz@tum.de}, Telephone: \,+49\,89\,289\,28429, Facsimile: +49\,89\,289\,28335}, N.C.~Truong, M.G.~Gerhofer, T.~Weyh, and H.-G.~Herzog \\
Institute of Energy Conversion\\
Technische Universit\"at M\"unchen, 80333 Munich, Germany
}

\maketitle

\begin{abstract}
Magnetic stimulation is a standard tool in brain research and many fields of neurology, as well as psychiatry. From a physical perspective, one key aspect of this method is the inefficiency of available setups. Whereas the spatial field properties have been studied rather intensively with coil designs, the dynamics have been neglected almost completely for a long time. Instead, the devices and their technology defined the waveform.

Here, an analysis of the waveform space is performed. Based on these data, an appropriate optimisation approach is outlined which makes use of a modern nonlinear axon description of a mammalian motor nerve. The approach is based on a hybrid global-local method; different coordinate systems for describing the continuous waveforms in a limited parameter space are defined for sufficient stability. The results of the numeric setup suggest that there is plenty of room for waveforms with higher efficiency than the traditional shapes. One class of such pulses is analysed further. Although the voltage profile of these waveforms is almost rectangular, the current shape presents distinct characteristics, such as a first phase which precedes the main pulse and decreases the losses. The single representatives, which differ in their maximum voltage shape, are linked by a nonlinear transformation. The main phase, however, seems to scale in time only.

\end{abstract}







\section{Introduction}

Magnetic stimulation is a standard tool for noninvasive activation of neurons, especially in the brain \cite{Weber:2002}.
The underlying physical principle of this method is induction, which is independent from any physical contact. Due to the magnetically passive behaviour of biological tissue, the magnetic field of the stimulation coil can pervade the body and especially poorly-conducting structures, such as bones. For a supra-threshold stimulation of brain neurons using electric currents, in contrast, the skull acts as a barrier. The consequently higher currents, however, lead to distress which is usually not tolerated by patients without anaesthesia \cite{Pechstein:1996}. Moreover, the current paths are no longer obvious, whereas inductive stimulation provides a relatively focused method for local stimulation of small neuron populations.

However, magnetic stimulation is extremely inefficient. Less than one percent of the electrical energy is transferred to the target. Although a substantial fraction of the field energy can be recuperated and used in the next pulse, the high currents in the range of kiloamperes during operation lead to significant ohmic losses in cables, connectors, and the stimulation coil itself. Driving such high charge flows necessitates up to 4000\,{}V. The application of high voltage is a serious issue for patient safety; the high losses heat the coil and limit typical durations of repetitive stimulation sessions to several minutes. Mobile devices with battery-powered pulse sources were proposed, but are not reasonable in the context of repetitive protocols.

Stimulation of neurons comprises two important domains. The current of the stimulation coil induces an electric field in the tissue, which in turn leads to drift currents. The role of induction with respect to efficiency was already discussed before \cite{Hsu:2003, Barker:1991a, Barker:1991b}. Two aspects lead to the small energy transfer in induction. The low conductance of biological tissue, which is more than six orders of magnitude lower than in copper, counteracts higher eddy currents. The second issue is the coupling between the primary windings---the stimulation coil here---and the tissue or a distinct region thereof, which corresponds to the secondary side of this \textit{transformer}. Improving the coupling is principally a geometric task, which may incorporate magnetic materials \cite{Epstein:2002}. This spatial degree of freedom can be covered with appropriate coil designs \cite{Ueno:1988, Lin:2000, Ruohonen:1997, Salinas:2007, Cadwell:1991, Jalinous:1991}.

The key to the temporal perspective is the second stage, namely the physiology of the neurons. Whereas the induction process is approximately linear, the neuron dynamics introduce a complex nonlinear system that may not be neglected in this context.
In the history of magnetic stimulation devices, efficiency has played a secondary role for a long time. First systems primarily aimed at reaching the stimulation threshold noninvasively with the available technological means---within more than a hundred years since J. C. Maxwell, many approaches have failed until this goal was finally reached \cite{Hallgren:1973, Barker:1982, Barker:1985, Geddes:1991}.

In the first devices, the entire energy was lost during a pulse. This so-called monophasic topology had been the standard for many years, until the biphasic oscillator circuitry introduced some improvements due to omitting the intentional damping. These were driven by purely technical considerations and allowed feeding back a notable fraction of the magnetic field energy into the pulse capacitor. The \lq{}needs\rq{} of the neurons were not taken into consideration also in the later device generations. Originally, it was thought that the falling phase of such an oscillation might counteract the activation of the \textit{leaky} neuron membrane \cite{Barker:1991b, Corthout:2001}. However, the system turned out to be not only rather efficient, but even more effective for the same voltage amplitudes \cite{Cadwell:1991, WeyhDiss, Niehaus:2000, Sommer:2002, Sommer:2006, Kammer:2001, Arai:2005}.

The question of dynamics is already an older issue in electrical stimulation. A lot of parameter studies were performed with certain predefined waveforms \cite{DeBock:2009, Grill:1995, Fang:1991, Wongsarnpigoon:2010a, Grill:1997, Dean:1985, Gorman:1983, Buetikofer:1978, Buetikofer:1979}. An analytic optimisation study, which led to the so-called rising exponential pulse, was reported already 1946 based on a variational approach \cite{Offner:1946}. A lot of work in this field has been done with Lapicque's well-known linear model from 1907 \cite{Lapicque:1907}. However, shortly after Offner's optimisation approach, it was shown that neuron dynamics are based on the interplay of a vast number of single components which are highly nonlinear \cite{HH:1952}. Despite that, his work with linear dynamics was reproduced over sixty years later \cite{Jezernik:2005}, but the results are neither compatible with more sophisticated neuron models \cite{Foutz:2010} nor with experiments \cite{Wongsarnpigoon:2010a}.

Nonlinear models, in contrast, can no longer be inverted easily, and chaotic behaviour of the corresponding differential equations render also the numerical handling problematic. A first general, (nearly) unbiased optimisation of the waveform for electrical stimulation was done recently \cite{Wongsarnpigoon:2010b}. The authors minimise the ohmic losses of monophasic electrical pulses. The optima reported there seem to be far more likely Gaussian than rising exponential.

For magnetic stimulation, there are only some rare parameter studies available in the literature related to different waveforms using experiments \cite{Havel:1997, Arai:2005, Claus:1990, Sommer:2006} and numerical models \cite{Hiwaki:1991, Reilly:1989, Nagarajan:1993, Maccabee:1998}.
Thus, even the knowledge about the different characteristics of existing waveforms is limited at the moment.

The question of optimality has not been studied so far. Instead of discussing again any technological changes of the device and their effects on neurons, this text wants to change the perspective and tries to address the requirements of a neuron instead, without any unnecessary restrictions, using a more realistic model.

\section{Setup}
\subsection{Objectives/Aims}

The central aim of this approach is a general optimisation without unnecessary constraints. Thus, no limitations of existing devices are taken into account. Only physical principles are of importance here.

For the objective function, several aspects of magnetic stimulation can be used. Here, a key problem of the technology is minimised, namely the energy losses. The heating of the stimulation coil limits the maximum duration of a session, as well as the repetition rate \cite{Weyh:2005}. Furthermore, the losses have a direct impact on the power consumption. The fact that portable repetitive devices are still missing is a consequence of that.

If no specific losses of a specific technology, such as the forward bias of p-n junctions in a concrete circuitry, are regarded, the main cause of power drain that can hardly be avoided is Joule heating relating to the inner resistance of all components. For a coil current $i(t)$, the losses of a pulse---regardless of the exact value of the inner resistance $R_i \neq 0$---are $\int_{\mathbb{R}^+} R_i i^2 \, dt \propto \int_{\mathbb{R}^+} i^2\, dt =: \mathcal{W}$. This integral forms the objective functional $\mathcal{W}$ for the further analysis.

In addition to the objective, several constraints are of importance---even without the limitations of a certain technology. Firstly, the coil current has to elicit a neuron response.
A second constraint results from the interplay of the objective and the response condition. As will become apparent later---and is known to many device designers---the losses at the excitation threshold fall with shorter pulse durations for classical waveforms. At the same time, the required device voltage increases \cite{Reilly:1989}.
In a consequent optimisation approach which does not regard that, the voltage diverges consequently. This degree of freedom can be determined by several means from a physical perspective. A simple voltage limit $v_{max}$ which may not be exceeded by the coil voltage $|v_c|<v_{max}$ in positive as well as in negative polarity was chosen here. Although this is often motivated by the limitations of semiconductor devices---regardless of the exact technology---it might be most adequate also with respect to patient safety and insulation. For being independent from the exact inner resistance, the voltage $v_c$ relates to the voltage of the inductance of the coil only. The current profile, in contrast, is independent from the losses.

\begin{figure}[t]
\centering
\includegraphics[width=1\columnwidth]{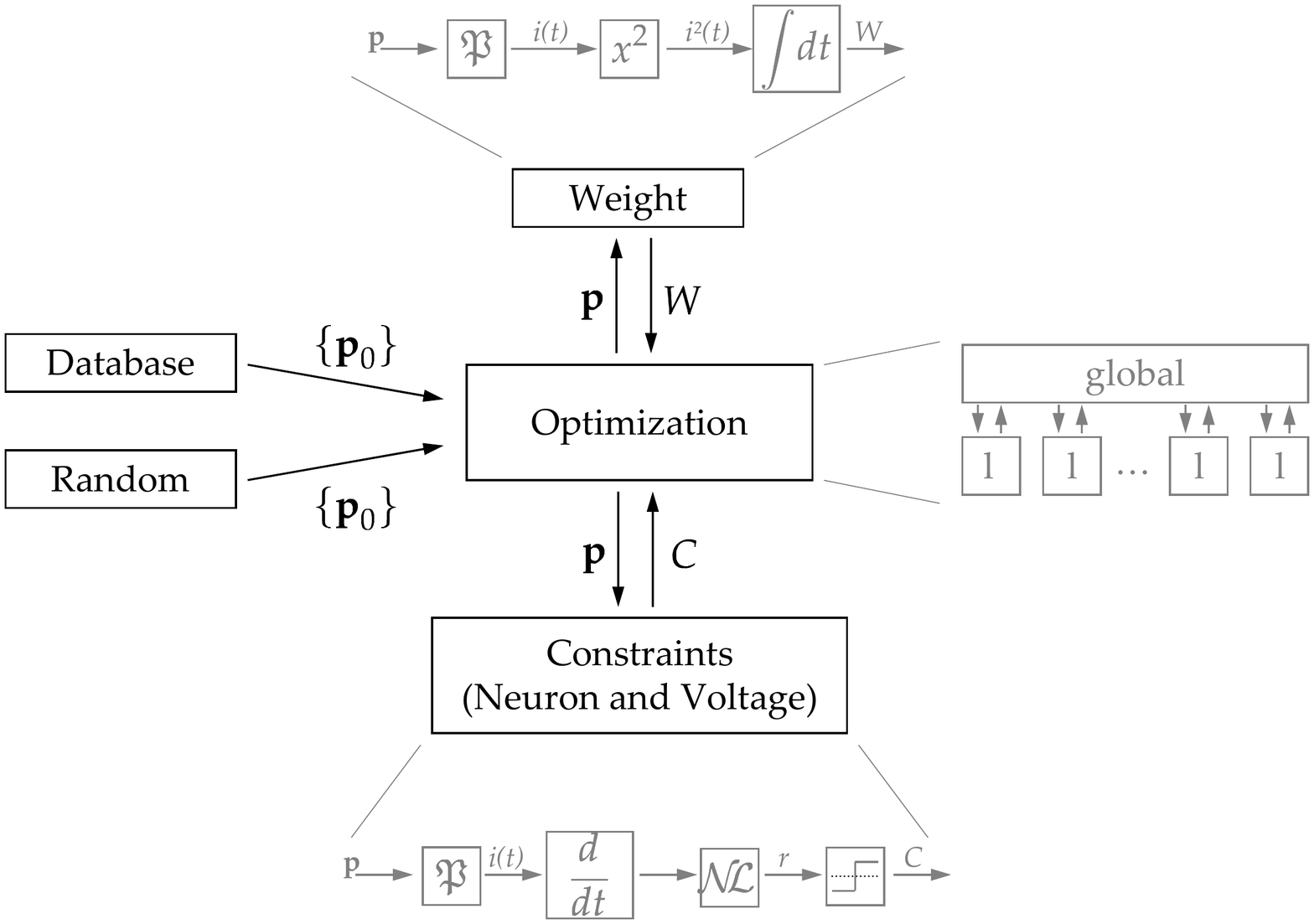}
\caption{\label{fig:StructureFull} Structure of the optimisation setup. The minimisation system itself has a hybrid structure. A global framework commands many local workers. The algorithms minimise the weight $W$ in the space of valid constraints $C$, but has only access to a finite number of abstract parameters ${\bf p}$. The latter are converted to functions by $\mathfrak{P}$ as described. The initialisation is performed with noise, as well as results from earlier runs. The axon model is incorporated into the constraints as a nonlinear element ($\mathcal{N\!L}$).}
\end{figure}

\subsection{Framework}

The optimisation framework has to describe the temporal behaviour, including all stages from the coil through to the response.
Figure \ref{fig:StructureFull} depicts the structure of the single modules.
The setup is outlined shortly as far as this information supports the understanding of the results.

The coil current induces an electric field. From a dynamic perspective, induction is a differentiating process. It is assumed that the tissue around the axons does not contain too many very different material types so that the filtering and distorting influence is rather low, and the first-order derivative becomes dominating. The latter provides the excitation function for the axon model. Further terms for describing distortion effects are optional, but set to 0 here. The neuron model, in turn, hands back the first constraint.

The second constraint and the objective function relate to the device as mentioned. Their evaluation is performed as described dependent on the voltage $v_c$ and the current $i$.

\subsection{Parametrisation/Problem Description}

The objective and all constraints were formulated as functionals. For numerical optimisation, however, this abstract construction has to be translated into a set of numbers. Thus, an appropriate definition of a parametrisation is an essential element, which moreover has a notable influence on the stability of the optimisation system. This mapping from the finite parameter space to a subspace of the Hilbert space of all waveforms allows the stable evaluation of all needed operators, such as derivatives, in an analytical way.
Evading this issue by taking a large number of sample points which are handed over to an optimisation algorithm \cite{Wongsarnpigoon:2010b} might end up in a lengthy \textit{brute-force} approach and instabilities.

The outlined conditions of magnetic stimulation are rather complex. Especially induction with its differentiating behaviour intensifies the problem. Lots of different waveforms are able to stimulate a nerve---their common features, however, are hardly recognizable. The part of the surface which is formed by the objective and furthermore fulfils the first constraint shows a strong tendency to a very rugged \lq{}crater landscape\rq{}.

The selection of an appropriate coordinate system for the description of the waveforms can remarkably improve the solution process, but the problem is nonlinear and has no natural (discrete) parametrisation. Three simple coordinate systems were used simultaneously here.

Firstly, the waveform is described by cubic spline curves; their parameters ${\bf p} \in \mathbb{R}^n$ act as degrees of freedom for the optimisation algorithm. Accordingly, this approach is similar to the core principles of the finite element method in numerics. The predefined solution forms a subspace of the class of all functions with continuous second derivative $C^2(T)$ with $T$ being the limited time domain of the reduced problem. However, in view of inductive stimulation, this coordinate system still has lots of unfavourable local minima in the objective function.

A parameter description in the frequency space was taken as an alternative. A complex-valued parameter vector ${\bf p'} \in \mathbb{C}^{{n}/{2}}$ can be taken without further constraints. A discrete Hilbert transformation provides the parameters for the single terms of a (complex-valued) Fourier series. Accordingly, the latter is even a \textit{smooth function} and a member of $C^\infty$.

Because the here applied optimisation algorithms work with simple floating-point numbers only, there are two ways for generating the complex-valued parameter vector ${\bf p'}$ from a real-valued vector ${\bf p} \in \mathbb{R}^n$ according to standard mathematics. On the one hand, the vector can be split into a real and an imaginary part. On the other hand, a partition into the absolute value and the phase is possible.\footnote{In case of the second method, the phase values can be increased by a constant factor in order to match their magnitude with the absolute values. This is an important issue for simplex-based optimisation methods as were used here. Those estimate the deviation usually from the next minimum on the basis of the area of a simplex. In the case of COBYLA, the \textit{goodness} of the simplex additionally decides over the type of the next step.}

All of the three proposed methods of parametrisation have at least two continuous derivatives. All of them allow a dynamic change of the coordinate system---this comprises a conversion from one parametrisation to another, as well as increasing or reducing the degrees of freedom in the same type of description.
All of them were incorporated here.

This may look like indecisiveness, but it was done intentionally in order to address the main problem of this endeavour, namely instability due to a vast number of local minima. A local minimum in one space might not be so pronounced in another. Lots of observed local minima showed artefacts typical for one type of description, such as Gibbs phenomena or \textit{ringing}, which are not stable for the optimiser with another parametrisation.
A typical example for a whole class of waveforms are the harmonic functions. A sinusoidal current stimulates as does its negative, i.e. mirrored version. In the time-domain, the shortest transition in-between passes 0, which does not stimulate at all and separates two minima. In certain frequency-domain descriptions, this is a less-problematic continuous shift of the phase parameters without such barriers.
Furthermore, a dynamic change of the number of dimensions was used for an adaptive control over stability. For a refinement, the degrees of freedom were increased. A diverging run can be re-stabilised by reducing complexity.

The parameters describe the current shape. Although a parametrisation of the electric field could avoid the differentiation of the function---integration is required instead---the results did not improve in that case.
The duration of the support of the waveforms was reduced to three milliseconds during the piloting phase of the setup as no local minimum was observed to require a longer time base.

\subsection{Optimisation Algorithm}

It seems a noteworthy hint in this context that not the objective, which is just squared, is the most challenging functional here, but the first constraint.

An optimisation method for multivariate minimization of the parametrised objective was implemented.
Due to the high number of local minima, a global method is advisable.
For a quick convergence, this was combined with a local algorithm in a hybrid approach.

Local optimisation workers with their much faster convergence run into the next local minimum. A global framework in turn combines the information about local optima.

For the global algorithm, a particle swarm was selected \cite{Eberhart:1995}, which seems to be more systematic in this context than defining genes for certain combinatory operations in a genetic algorithm, for instance. In addition to that, a particle-swarm framework with lots of local workers very easily allows a concurrent design that meets the requirements of high-performance computing.

Two algorithms---a simplex descendent (constrained optimisation by linear approximation, COBYLA, \cite{Powell:2003}) and an interior-point method \cite{Waltz:2006}---act alternatively as relatively stable local optimisers. The type of the coordinate system of a certain worker is fixed. The number of degrees of freedom applies to all workers and is controlled by the global framework. The number is increased by a predefined step if convergence is achieved and the result outperforms the latter; otherwise, it decreases.

The global method hands over the parameters to the local workers, which are supposed to run into a nearby valid local minimum which fulfils all constraints. Their results are regarded for deciding on the particles' as well as the total best in each step. In order to avoid oscillations in the attraction field of local minima, those are not assigned to the current position of a particle.

The update rules of the particle swarm assign the following to the parameters of the $j$-th particle in the $(i+1)$-th step
\begin{eqnarray}
{\bf p}_j^{i+1} &=& {\bf p}_j^i + {\bf \Delta p}_j^{i+1}		\\
{\bf \Delta p}_j^{i+1} &=& \omega {\bf \Delta p}_j^i + c_1 r_{j1} ({\bf b}_j - {\bf p}_j^i) + c_j r_{j2} ({\bf b}_g - {\bf p}_j^i),
\end{eqnarray}
with the inertia $\omega$, the gravity parameters $c_1$, $c_2$ of the local and the global best, as well as the modulation variables $r_{j1}, r_{j2} \in (0, 1)$. The latter are chosen randomly in every step. The global best is denoted by ${\bf b}_g$, whereas each particle has its own local best ${\bf b}_j$. For the constants, several alternatives were tested. The following values performed appropriately: $\omega = 0.6$, $c_1 = 1.7$, $c_2 = 1.7$, alternatively even with repellent behaviour according to $\omega = -0.35$, $c_1 = -0.05$, $c_2 = 5$. The number of particles was up to fifty.

Initialization of the single particles is done with random numbers. In several runs, a fraction of the workers was initialized with the known classical waveforms.

\subsection{Neuron Model}

The decision if a pulse elicits an action potential is performed by a nonlinear model of a human motor axon. The local description (see the appendix), which is an adaptation from the literature, includes fast sodium channels, persistent sodium channels, one type of potassium channels and passive leakage assembled on the basis of \cite{McIntyre:2002, Reid:1999, Schwarz:1995, Scholz:1993, Safronov:1993}. Its consistency with stimulation experiments which address especially the nonlinear behaviour of nerves was studied in \cite{GoetzDiss}.
A response that exceeds $+10$\,{}mV is seen as successful.

The whole implementation is done in C and optimised for maximum speed. For the solution of the differential equation, a standard second-order Runge-Kutta method was used. The allowed maximum time step for the final analysis was set to $500$\,{}ns. The whole optimisation framework followed a strict parallel design. Every call of the differential-equation solver is executed in an own thread. The computation was performed on three Xeon servers with eight cores in each for the pre-evaluation and tests, as well as on a high-performance system of the Leibniz Supercomputing Centre of the Bavarian Academy of Sciences and Humanities. The results of this paper are based on more than $50\,{}000$ CPU hours.

\begin{figure}[t]
	\centering
		\includegraphics[width=0.95\columnwidth]{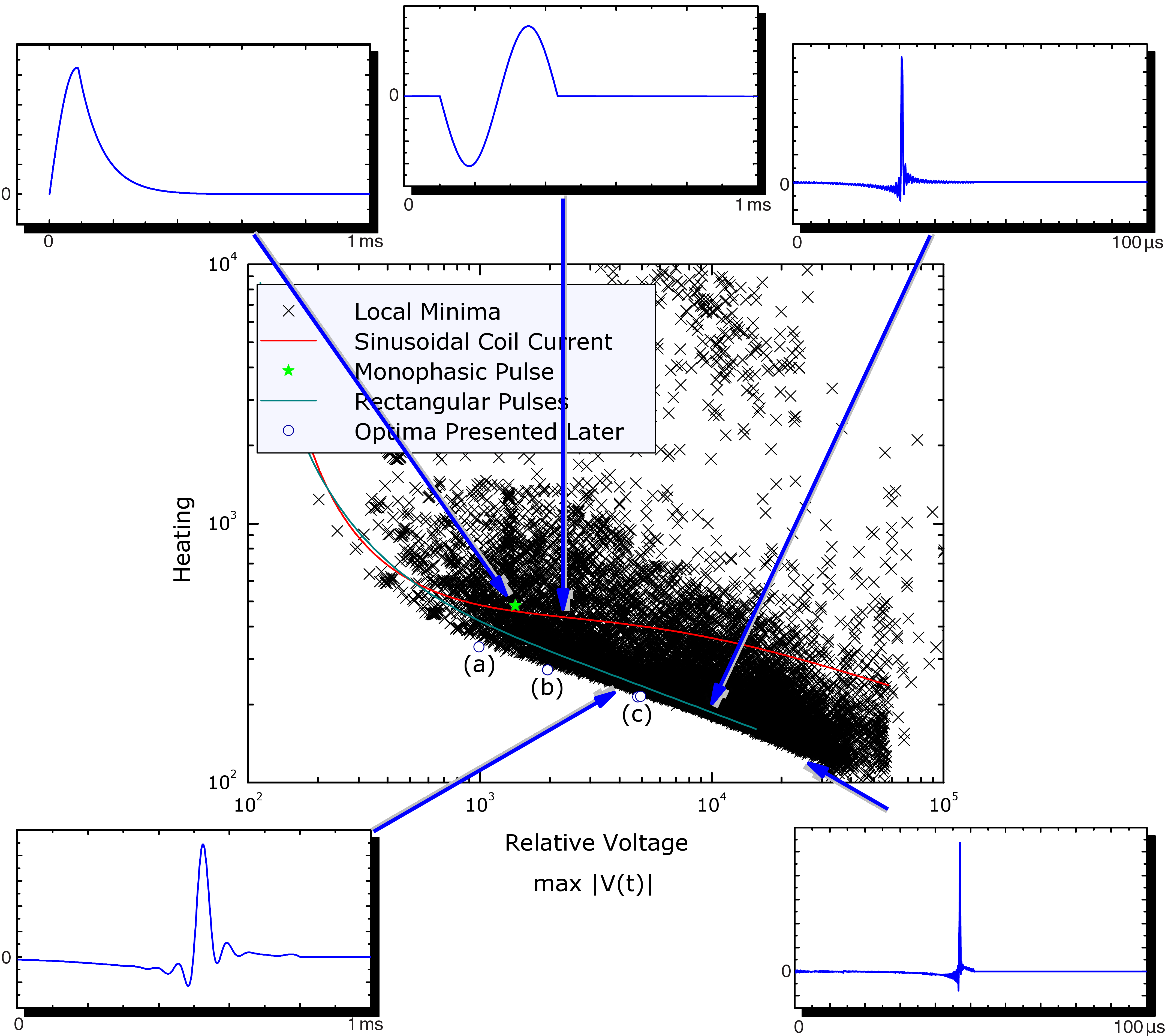}
	\caption{\label{fig:optimizationsinglebee01} Local minima from the pre-evaluation plotted in a plane which is spanned by the peak voltage and the pulse losses at the excitation threshold. Additionally, biphasic and rectangular (with respect to the voltage) pulses were depicted for a wide range of carrier frequencies. The insets outline the current shapes some examples. The small letters refer to solutions which are shown in figure \ref{fig:Optima1000DOFcomparison01}.}
\end{figure}

\section{Pre-Evaluation}

The study of different waveforms from a physical perspective has been neglected so far in inductive stimulation. Even basic relationships are not generally known, but are mostly specific experience of academic and commercial device designers. Compared to electrical stimulation, the circumstances are more complex because of induction. Therefore, they are also less obvious.

For that reason, a pre-evaluation of solutions was performed. Additionally, this acted also as a test of the stability of the algorithms at the same time. The results are presented in figure \ref{fig:optimizationsinglebee01}. This graph depicts the losses in dependence of the maximum required voltage at the threshold. All values are relative because the threshold is nevertheless an individual figure. However, the voltage is roughly in the range of real devices if the unit volt is added. Exact calibration can be conducted easily based on the data.

Every cross in the graph represents a local optimum. In addition to them, lines for the standard biphasic pulse (red line) and rectangular voltage pulses (cyan, dashed) were added from a parameter study for a frequency range from $500$\,{}Hz to $50$\,{}kHz. Moreover, the waveform of a commercially-available monophasic device (Magstim 200, Magstim Ltd., Whitland, Wales) was added (star). For all of them, that current direction was taken into account which presented the lowest threshold.

Most interesting in this graph is that the monophasic waveform is not notably less efficient than a biphasic pulse in the model. It seems to be the way of generation with the intentional damping only that wastes the pulse energy. This is remarkable because the current tail is rather long so that it makes a substantial contribution to the loss integral. This might call for better topologies since monophasic pulses have several distinct features, such as the higher sensitivity to the coil orientation, which is favourable in a lot of applications \cite{Niehaus:2000, BrasilNeto:1992}.

Nearly all waveform types seem to have the general tendency that the losses can be reduced if a higher voltage is provided in the device. This was already mentioned in the text above. For biphasic and rectangular stimuli, the transformation is just a simple dilation or compression. Although this effect is sometimes used for reducing the heating in high-power applications, e.g. in rehabilitation \cite{Szecsi:2010}, there seem to be pulse shapes that have a much steeper slope in this loss-voltage plane, i.e. the reduction of the losses for an increased voltage is notably higher than for biphasic pulses.

Rectangular pulses might be more efficient than the classical waveforms. This assumption was already stated before \cite{Peterchev:2011, Havel:1997}. However, they are still far from being optimal---especially for the voltage range of typical devices.

The lower edge of the point cloud is smooth and seems to represent the same class of stimuli. For very high voltages, the current waveforms degenerate to Dirac-like pulses.

\begin{figure}[t]
	\centering
	
	\subfigure{\includegraphics[width=0.75\columnwidth]{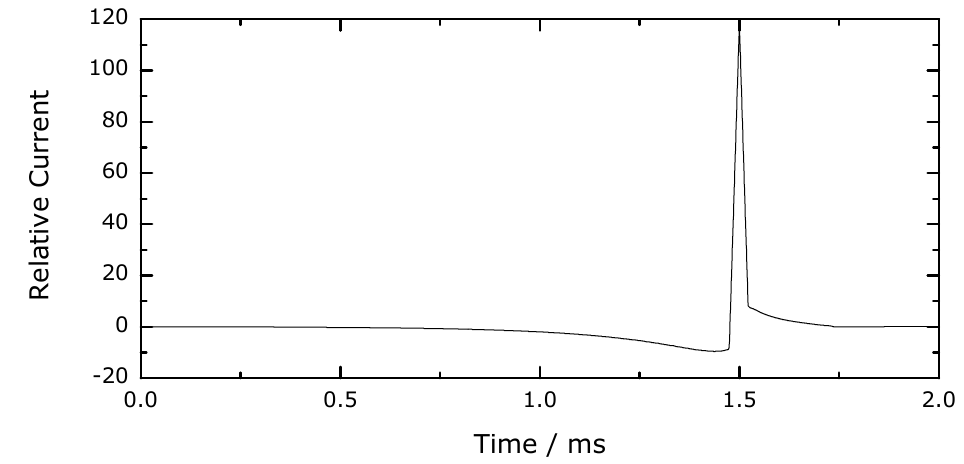}}
	\subfigure{\includegraphics[width=0.75\columnwidth]{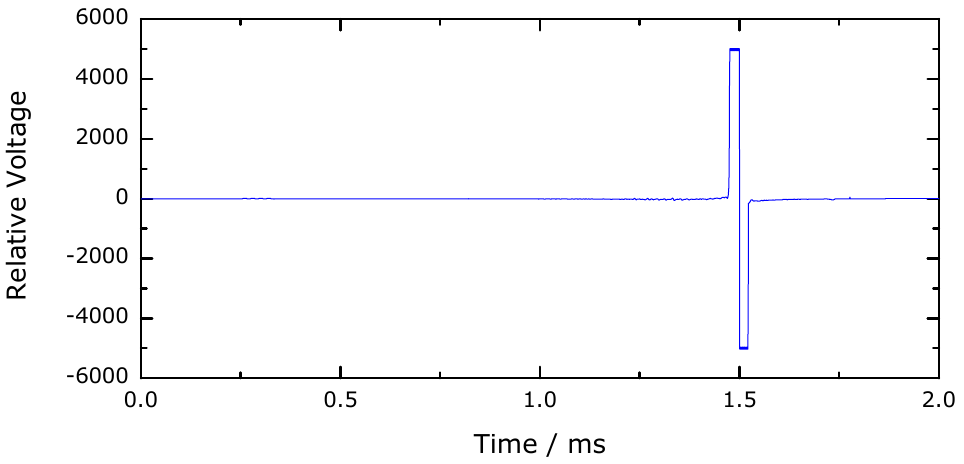}}
	
	\caption{\label{fig:Ref5000V_weight215} Relative current and voltage profiles for one exemplary minimum. The absolute values on the axis depend on many parameters, such as coil design, inductance, distance, etc. The full-width at half maximum of the main phase of the current pulse amounts to $23$\,\textmu{}s (see figure \ref{fig:optimizationsinglebee01}(c)). The maximum derivative within the first phase is by about a factor of $161$ lower than the slopes of the second. Despite the low amplitude, the long duration causes that the area under the curve for the first current phase is higher than that of the second (by $15$\,\%).}
\end{figure}

\section{Optimisation Results}
\subsection{Basic Features}

For several different voltages in the range of classical devices, a more exact analysis was performed including a computationally-intensive refinement. The number of degrees of freedom was increased up to one thousand during these runs. Most of the computational resources were spent on this investigation.
Some exemplary results are depicted in figures 3 and 4.

The best found solutions are very similar. The current always exhibits three relatively distinct phases. A slow first phase precedes a short main pulse in opposite direction; a separate third phase becomes visible after very high iteration numbers of the optimisation algorithms.

The shapes of the slopes within the single phases seem to have only minor influence on the losses. However, they define the maximum voltage level. The fine-analysis reveals an almost rectangular voltage shape of the second phase. Accordingly, the current is approximately triangular there. The other parts of the pulse seem to disappear in the voltage profile.

The voltage, however, cannot be the benchmark because the pre-evaluation clearly shows that rectangular pulses are definitely suboptimal in this model. It has merely indirect influence. The current, in contrast, shows rather subtle characteristics.

The square-shaped voltage during the main phase might be the consequence of the second constraint. The system uses the maximum allowed voltage level as long as possible. This explains the positive wing of the voltage. The negative, however, reduces the current as quickly as possible after the peak. Although this might counteract excitation, a lower current level---which has squared influence in the losses---might be more important due to the objective. This occurs only until a certain low level is reached from which the third phase heads for 0 in a slower, exponential decay.

This reasoning can also be used for making the first phase plausible. The latter biases the onset of the second phase with minimum dynamics. Starting from this shifted baseline, a rather long rising slope can emerge, without reaching extremely high loss powers.
Also this first phase causes losses, but the amplitude is relatively low. Since the heating depends on the squared current, this investment seems to pay off. Prolonging the slope to the left, thus to negative values, leads to much lower costs in the objective than increasing the peak current. For the waveform of figure \ref{fig:Ref5000V_weight215}, the losses would be approximately $11$\,\% higher without the first phase; for lower voltages this difference is much higher.\footnote{In the voltage range of the incorporated commercial monophasic device, the losses of the rectangular pulse seem to be even $25$\,\% higher than in the corresponding local minimum with the same voltage limits (see figure \ref{fig:RectVsLocalOptimOverDuration}).}

Nevertheless, it should be noted that such explanation is always an oversimplification. Ascribing certain functions to different phases of a pulse can be misleading in connection with nonlinear dynamics, although older literature might motivate such a step.

\begin{figure}[t]
	\centering
		\includegraphics[width=0.75\columnwidth]{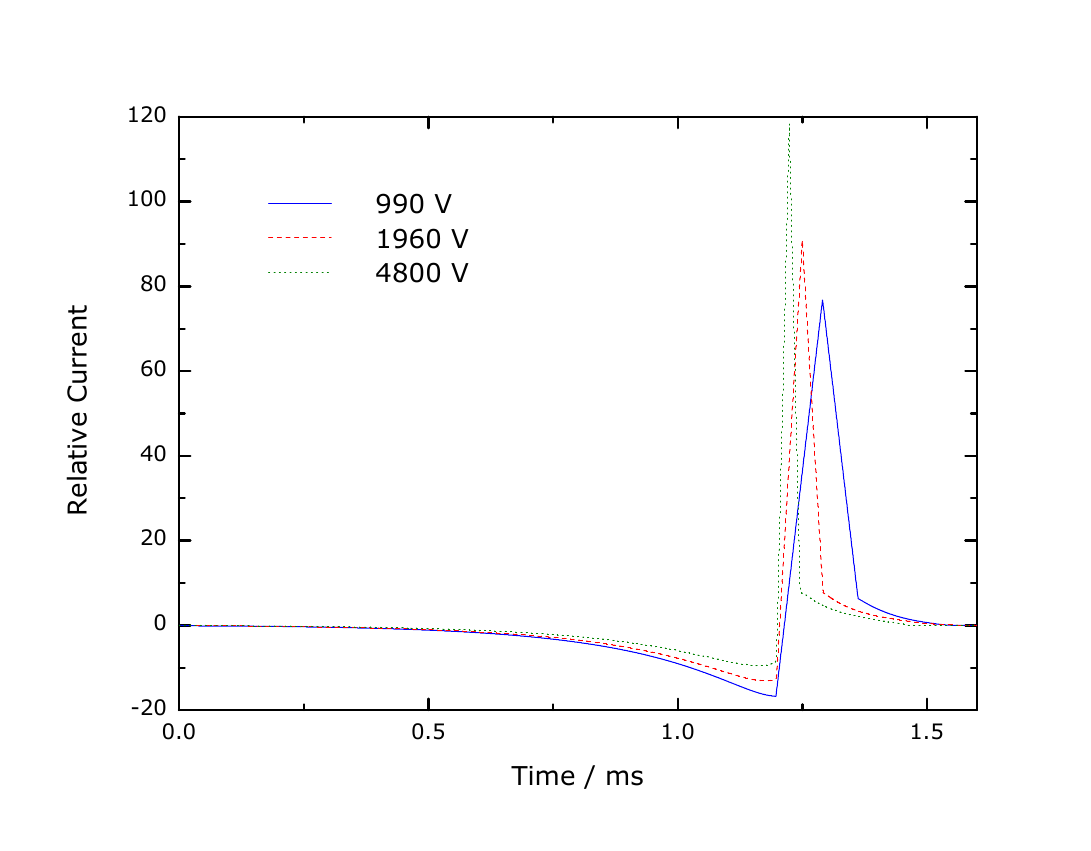}
	
	\caption{\label{fig:Optima1000DOFcomparison01} Local pulse optima for different threshold-voltage constraints.
	The comparison reveals the different roles of the single parts. Whereas the length of the second phase roughly scales inversely to the voltage constraint, the duration of the first part stays constant. However, the amplitude of the first phase relative to the second varies. The exact data in the coordinate system of figure \ref{fig:optimizationsinglebee01} are as follows: $334$ as the corresponding loss value for a relative voltage of $990$ (a), $272$ for a relative voltage of $1960$ (b), as well as a loss value of $214$ for a voltage of $4800$ (c).}
\end{figure}

\subsection{Common Ground of the Corresponding Class of Waveforms}

The single curves in figure \ref{fig:Optima1000DOFcomparison01} suggest that these waveforms can be treated as representatives of the same class. In contrast to biphasic or rectangular pulses, the common ground of the members is not a linear transformation, such as scaling. The behaviour of the first phase---which, for instance, even decreases if the main peak becomes stronger---speaks out against such a simplification.

If the second phase is isolated, however, a simple time scaling can be assumed in the first approximation. The pulse duration decides on the voltage level and influences the heating: For shorter triangles, the amplitude has to be increased, but the losses fall. The dynamics of the first phase, however, do not change at all for the studied range, but only the amplitude is adjusted for the different voltage constraints.

\begin{figure}[t]
	\centering
	
	\subfigure{\includegraphics[width=0.75\columnwidth]{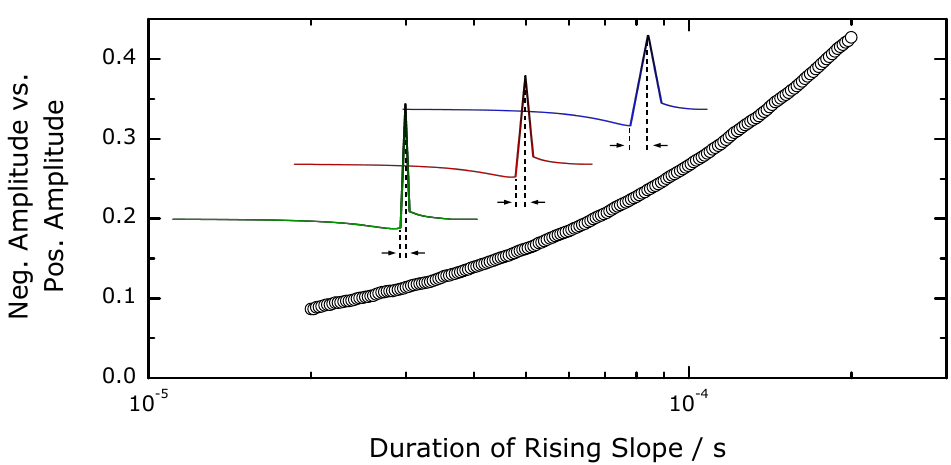}}
	\subfigure{\includegraphics[width=0.75\columnwidth]{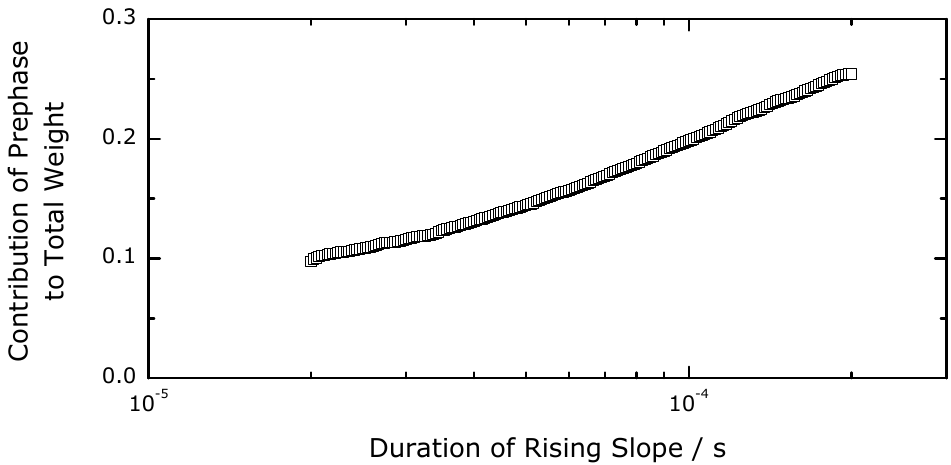}}
	
	\caption{\label{fig:ParameterstudyOptimumPrephaseAmplitude} Role of the first phase of the current for different representatives of the whole class of local minima. If the pulse duration is increased, the amplitude of the first phase rises relatively to the main pulse and leads to a nonlinear transformation rule between the pulses. The contribution of the first phase to {ohmic} losses is also not constant. The x-axis refers to the length of the rising slope. This is equivalent with the duration of the first voltage peak.}
\end{figure}

The exact behaviour of such details was analysed in a parameter study. The amplitude of the first phase, i.e. the value of the most negative point in the current profile, was rendered changeable in dependence of the main-phase peak amplitude. As a second degree of freedom, the duration of the rising slope of the main phase was varied; the falling slope changes proportionally to the latter. The optimal amplitude of the first phase was evaluated for a number of different main-pulse durations. The results are presented in figure \ref{fig:ParameterstudyOptimumPrephaseAmplitude}.

The curves depict the amplitude relation as well as the contribution of the first phase to the objective, thus the value of the integral of the squared current over the first phase in relation to the total weight. For very short pulses, the first part of the waveform vanishes almost and the pulse becomes nearly triangular in the current profile, as well as rectangular with regard to its voltage shape.
Also the investment in the prephase seems to be less rewarding. For the longest depicted main pulse durations, however, the current at the apex of the first phase is nearly half of the main amplitude.

\begin{figure}[t]
	\centering	\includegraphics[width=0.75\columnwidth]{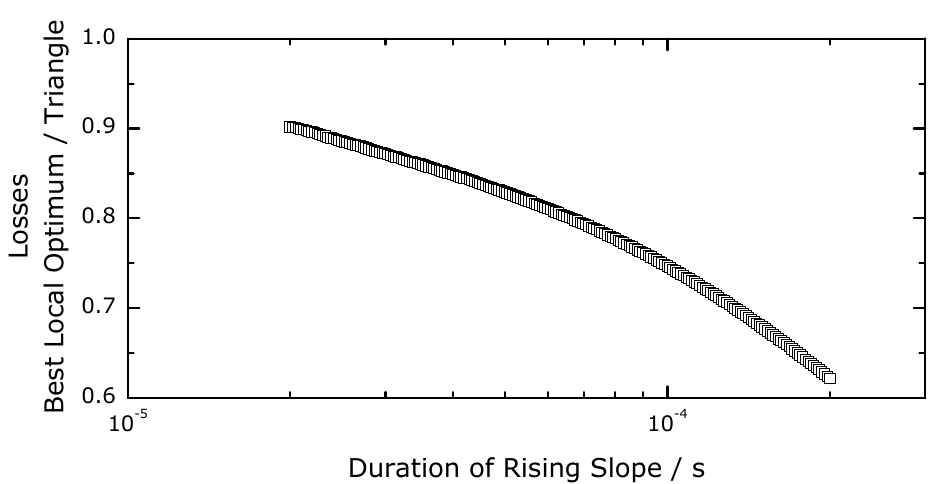}
	
	\caption{\label{fig:RectVsLocalOptimOverDuration} Influence of the pulse duration on the advantage of the local minima compared to simple triangular current pulses. The y-axis depicts the relation of the ohmic losses of the corresponding local optimum to the heating of the corresponding triangular current profile (hence rectangular voltage) with the same main pulse duration; both are at the excitation threshold. Even for unusually-short pulses with some tens of microseconds, the gain is still more than $10$\,\%.}
\end{figure}

Figure \ref{fig:RectVsLocalOptimOverDuration} shows the effect for different pulse durations in relation to symmetric triangular current profiles, i.e. rectangular voltage pulses, with the same duration. 
With the increasing amplitude of the first phase in pulses with a longer second part, the advantage over the simpler competitor increases. For a duration of the rising triangular slope of $200$\,\textmu{}s, the local minimum amounts to $62$\,\% of the losses for the corresponding triangular current shape only. Again, for very short pulses, the difference in weight between the local optima and the triangular current pulses vanishes because also the differences in respect of the shape fade away.

\begin{figure}[t]
	\centering
		\includegraphics[width=0.75\columnwidth]{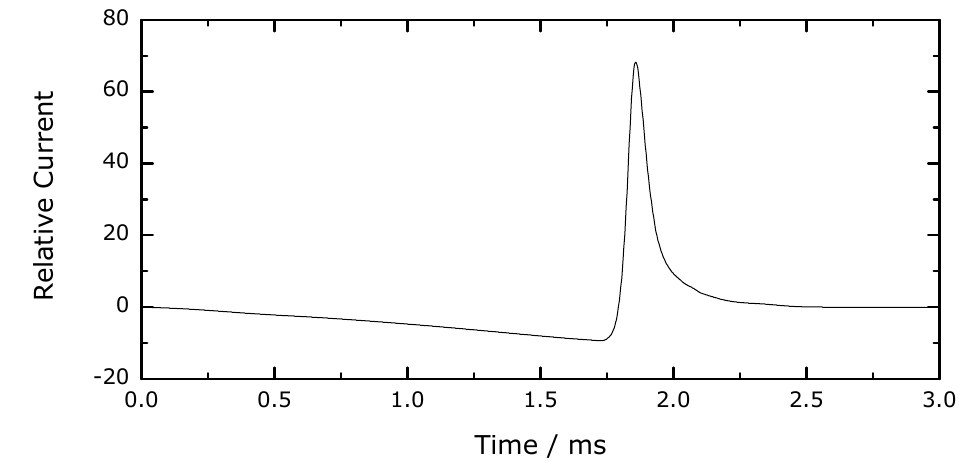}
	\caption{\label{fig:HHk12_optim_160DOF01} Local minimum from another axon model (Motz-Rattay). The essential elements from the more realistic model above appear also here.
}
\end{figure}

{
For the question of plausibility, a short look at corresponding results from another model might be worthwile. Taking the original Hodgkin-Huxley equations and modifying especially their dynamics to mammalian body temperatures once was a common simple and moreover computationally quick approach for studying pulse dynamics, e.g. of the acoustic nerves \cite{MotzRattay:1986}.
Although it is not appropriate for the analysis of different waveforms due to the notably different dynamics, the best found result from figure \ref{fig:HHk12_optim_160DOF01} also shows the known key phases.
}

\section{Conclusions}

In this text, a minimisation approach was formulated for inductive stimulation of an axon model; the key objective was the losses. Their minimum level is given by an ohmic linear voltage-current relationship, which leads to a squared dependence on the current for the heating.

The approach was not limited to a specific circuitry, but the constraints were reduced to the necessary conditions only.
The parametrised waveform space covers a very wide range of functions and is---from a practical perspective---almost unconfined.

Although many of the worse local minima that emerged during the computations presented rather curious pulse shapes, the best results found here seem very consistent. The single parts of the waveforms can be made plausible as shown, although this should not be misinterpreted as an explanation, which is a critical task in the context of nonlinear dynamics.

All waveforms were reported with their current profile. This was done in order to emphasise the different roles of voltage and current for the predefined objective. In the past, the voltage profile was usually foregrounded. Also in typical devices, the voltage level is controlled rather than the current \cite{Cadwell:1991, Jalinous:1991, Barker:1991a, McRobbie:1985, Peterchev:2011}. Concerning the losses, however, the current is the key figure. 
For the voltage profile, in contrast, the reference is not even well-defined. As the currents are relatively strong, the voltage drops at connectors, cables, etc.\ are high enough that the differences of the voltage shapes due to moving the monitoring points is notable.

The potential of waveforms---not only for optimisation, but also in the context of additional degrees of freedom---seems to have only secondary priority for commercial device manufacturers. The authors are not aware of any commercially-available system that would be able to provide the here derived shapes. This pulse generation has to be necessarily with recuperation of the field energy so as not to counteract the objective. Recently, a sophisticated technology was proposed by an academic team in order to provide more flexibility \cite{Peterchev:2011, Peterchev:2010}. Although the described embodiment, which was designed for relatively low negative voltages, is not fully compatible with the aims given here, the underlying principle of that system could provide similar waveforms, at least.

An alternative technology that would allow even a rather exact reproduction of such pulses was developed most recently \cite{GoetzDiss} and will be discussed separately.

From the perspective of efficiency, another tradition has to be reviewed. In classical devices, the amplitude is used for controlling the strength. Due to the \lq{}monochrome\rq{} oscillator design, an alternative was not possible before cTMS \cite{Peterchev:2011}. In the light of the falling losses for all classical pulse shapes, it could be advantageous in several high-power applications \cite{Szecsi:2010} to use the highest available voltage level of a device and modulate the strength with the pulse width instead. 
For sessions where the dynamics are important for certain effects, such as selective stimulation or diagnostics, the established amplitude control can be applied.

Although the results were obtained in a systematic way and based on a plausible explanation, statements derived from any type of model should in general not be taken as irrevocable facts, but are rather a suggestion for experiments.
The outcome was not biased by an anticipating initialisation with current or voltage shapes that resemble the final outcome, but the waveforms evolved themselves. Many artefacts and worse local minima show much different curves.

Despite that, the results are only (the best found) local minima. Moreover, they are relatively stable and re-emerged also in totally independent runs with random initialisation of the parameters. Indeed, these solutions can be global or not far from the global solution, but they do not have to.

Furthermore, the found waveforms are undeniable local minima of the problem formulation, which is based on a nonlinear axon model. Despite calibration and tests, each model shows limitations, shortcomings, and does only mimic reality. Especially in the absolute task of minimization, a physiologic optimum could look slightly different, and even be an individual quality of a nerve.
Thus, this is also a request to experimentalists to further study this issue.

\section*{Acknowledgement}
The Leibniz Supercomputing Centre of the Bavarian Academy of Sciences and Humanities (\textit{www.lrz.de}) provided computational resources for supporting this project.

\section*{Author Contributions}

SMG and TW designed the research; SMG devised the theoretical and mathematical foundation. SMG, MGG, and NCT implemented the code. MGG and NCT tuned the optimisation methods and managed the super-computing facilities. All authors critically reviewed the data and discussed the approaches. SMG wrote the manuscript. MGG and NCT contributed equally to this work.


\pagebreak
\section*{Appendix}

The state vector $\phi$ of the nerve model is defined as follows. The first element denotes the membrane potential; the remaining items describe the channel dynamics, which are represented by first-order systems with non-linear time constants:

\setlength{\parindent}{0mm}

\begin{eqnarray}
\phi = \left( \begin{array}{c} \phi_1\nonumber\\ \phi_2\nonumber\\ \phi_3\nonumber\\ \phi_4\nonumber\\ \phi_5	\end{array}\right) \equiv \left(\begin{array}{c} v\nonumber\\ m\nonumber\\ h\nonumber\\ p\nonumber\\ s	\end{array}\right)
\end{eqnarray}

\begin{equation}
	\frac{\partial}{\partial t} \phi	=	\left(\begin{array}{c} \frac{1}{c_m}\left(i_{f} + \left(g_{Na} \phi_2^3 \phi_3 +  g_{Na,p} \phi_4^3\right)(V-E_{Na}) - g_K \phi_5 (V-E_K) - g_l (V-E_l)\right)\nonumber\\ \frac{m_\infty - \phi_2}{\tau_m(\phi_1)}\nonumber\\	\frac{h_\infty - \phi_3}{\tau_h(\phi_1)}\nonumber\\	\frac{{p_\infty} - \phi_4}{\tau_{p}(\phi_1)}\nonumber\\	\frac{s_\infty - \phi_5}{\tau_s(\phi_1)} \end{array}\right)
\end{equation}

\vspace{12pt}

\begin{multicols}{2}

  $c_m = 2.0$\,\textmu{}F/cm$^2$	\\
	
	$g_{Na}	= 290$\,{}mS/cm$^2$	\\
	$g_{Na,p} = 25$\,{}mS/cm$^2$	\\
	$g_{K} = 80$\,{}mS/cm$^2$	\\	
	$g_l = 7$\,{}mS/cm$^2$	\\
	$E_{Na} = 50.0$\,{}mV	\\	
	$E_{K} = -90.0$\,{}mV	\\
	$E_l	= -90.0$\,{}mV	\\
$T = 310$\,{}K\\

	$\delta_m = 1.86$\,{}(mV\,{}ms)$^{-1}$	\\
	$\epsilon_m = 21.4$\,{}mV	\\
	$\zeta_m = 10.3$\,{}mV	\\
	$\eta_m = 0.086$\,{}(mV\,{}ms)$^{-1}$	\\
	$\theta_m = 25.7$\,{}mV	\\
	$\iota_m = 9.16$\,{}mV	\\
	$\delta_h = 0.062$\,{}(mV\,{}ms)$^{-1}$	\\
	$\epsilon_h = 114.0$\,{}mV	\\
	$\zeta_h = 11.0$\,{}mV	\\
	$\eta_h = 2.3$\,{}ms$^{-1}$	\\
	$\theta_h = 31.8$\,{}mV	\\
	$\iota_h = 13.4$\,{}mV	\\
	$\delta_{p} = 0.01$\,{}(mV\,{}ms)$^{-1}$	\\
	$\epsilon_{p} = 27$\,{}mV	\\
	$\zeta_{p} = 10.2$\,{}mV	\\
	$\eta_{p} = 2.5 \cdot 10^{-4}$\,{}(mV\,{}ms)$^{-1}$	\\
	$\theta_{p} = 34$\,{}mV	\\
	$\iota_{p} = 10$\,{}mV	\\
	$\delta_s = 0.3$\,{}ms$^{-1}$	\\
	$\epsilon_s = 53$\,{}mV	\\
	$\zeta_s = -5$\,{}mV	\\
	$\eta_s = 0.03$\,{}ms$^{-1}$	\\
	$\theta_s = 90$\,{}mV	\\
	$\iota_s = -1$\,{}mV	\\

	$T_{r,act} = 293$\,{}K	\\
	$T_{r,dea} = 293$\,{}K	\\
	$T_{r,K} = 309$\,{}K	\\
	
	$\kappa_{act} = \exp\left(\frac{(T-T_{r,act})\ln{2.2}}{10\,\textrm{K}}\right)$	\\
	$\kappa_{dea} = \exp\left(\frac{(T-T_{r,dea})\ln{2.9}}{10\,\textrm{K}}\right)$	\\
	$\kappa_{K} = \exp\left(\frac{(T-T_{r,K})\ln{3.0}}{10\,\textrm{K}}\right)$	\\

	$\alpha_m = \kappa_{act} \psi_{m \alpha}(\phi_1)$	\\
	$\beta_m = \kappa_{act} \psi_{m \beta}(\phi_1)$	\\
	$\tau_m = \frac{1}{\alpha_m + \beta_m}$	\\
	$m_\infty = \frac{\alpha_m}{\alpha_m + \beta_m}$	\\

	$\alpha_h = \kappa_{dea} \psi_{h \alpha}(\phi_1)$	\\
	$\beta_h =  \frac{\kappa_{dea} \eta_h}{1 + \exp\left(-\frac{v+\theta_h}{\iota_h}\right)}$	\\
	$\tau_h = \frac{1}{\alpha_h + \beta_h}$	\\
	$h_\infty = \frac{\alpha_h}{(\alpha_h + \beta_h)}$	\\

	$\alpha_p = \kappa_{act} \psi_{p \alpha}(\phi_1)$	\\
	$\beta_p = \kappa_{act} \psi_{p \beta}(\phi_1)$	\\
	$\tau_p = \frac{1}{\alpha_p + \beta_p}$	\\
	$p_\infty = \frac{\alpha_p}{\alpha_p + \beta_p}$	\\

	$\alpha_s = \frac{\kappa_{K} \delta_s}{1 + \exp\left(\frac{\phi_1 + \epsilon_s}{\zeta_s}\right)}$	\\
	$\beta_s = \frac{\kappa_{K} \eta_s}{1 + \exp\left(\frac{\phi_1 + \theta_s}{\iota_s}\right)}$	\\
                
	$\tau_s = \frac{1}{\alpha_s + \beta_s}$	\\
	$s_\infty = \frac{\alpha_s}{\alpha_s + \beta_s}$	\\

	$\psi_{m \alpha} = \left\{ \begin{array}{ll} \delta_m \zeta_m	&	\textrm{for}~\phi_1 = - \epsilon_m	\\
								\frac{\delta_m (\phi_1 + \epsilon_m)}{1 - \exp\left(-\frac{\phi_1 + \epsilon_m}{\zeta_m}\right)}	&	\textrm{else}	\end{array}\right.$	\\
	$\psi_{m \beta} = \left\{ \begin{array}{ll} - \eta_m \iota_m	&	\textrm{for}~\phi_1 = - \theta_m	\\
								-\frac{\eta_m (\phi_1 + \theta_m)}{1 - \exp\left(\frac{\phi_1 + \theta_m}{\iota_m}\right)}	&	\textrm{else}	\end{array}\right.$	\\
	$\psi_{h \alpha} = \left\{ \begin{array}{ll} - \delta_h \zeta_h	&	\textrm{for}~\phi_1 = - \epsilon_h	\\
								-\frac{\delta_h (\phi_1 + \epsilon_h)}{1 - \exp\left(\frac{\phi_1 + \epsilon_h}{\zeta_h}\right)}	&	\textrm{else}	\end{array}\right.$	\\
	$\psi_{p \alpha} = \left\{ \begin{array}{ll} \delta_{p} \zeta_{p}	&	\textrm{for}~\phi_1 = - \epsilon_{p}	\\
								\frac{\delta_{p} (\phi_1 + \epsilon_{p})}{1 - \exp\left(-\frac{\phi_1 + \epsilon_{p}}{\zeta_{p}}\right)}	&	\textrm{else}	\end{array}\right.$	\\
	$\psi_{p \beta} = \left\{ \begin{array}{ll} - \eta_{p} \iota_{p}	&	\textrm{for}~\phi_1 = - \theta_{p}	\\
								-\frac{\eta_{p} (\phi_1 + \theta_{p})}{1 - \exp\left(\frac{\phi_1 + \theta_{p}}{\iota_{p}}\right)}	&	\textrm{else}	\end{array}\right.$	\\

\end{multicols}

\end{document}